\documentstyle[aps,twocolumn,psfig]{revtex}
\newcommand{\bc}{\begin{center}}
\newcommand{\ec}{\end{center}}
\newcommand{\half}{\frac{1}{2}}
\newcommand{\be}{\begin{equation}}
\newcommand{\ee}{\end{equation}}
\newcommand{\bea}{\begin{eqnarray}}
\newcommand{\eea}{\end{eqnarray}}
\newcommand{\nn}{\nonumber}

\newcommand{\x}{\vec{x}}

\newcommand{\bphi}{\phi_{\rm cl}}
\newcommand{\bpi}{\pi_{\rm cl}}
\newcommand{\D}{\mbox{D}}

\begin{document}

\preprint{ITFA-98-20\\ Regensburg TFT98}

\title{{\small\begin{flushright}
ITFA--98--20 \\ hep--ph/9809378 
\end{flushright}}Real-Time Dimensional Reduction in Thermal Field Theory\thanks{Talk 
given at TFT98, Regensburg, Germany, August 10-14, 1998.}}

\author{Bert-Jan Nauta\thanks{mail: 
nauta@phys.uva.nl} and Chris van Weert}

\address{Institute for Theoretical Physics, University of Amsterdam\\
        Valckenierstraat 65, 1018 XE  Amsterdam, The Netherlands}

\date{\today}

\maketitle

\begin{abstract}
We consider the extension of static  dimensional reduction to real-time. For 
a scalar field theory it is shown that in the high-temperature limit this 
leads  to an effective classical theory.  
Quantum corrections  to the leading classical behavior are determined by an 
effective action which can be calculated perturbatively in the background of 
the classical fields.  Feynman rules for \(\lambda\phi^4\)-theory are given 
and  the extension to SU($N$) gauge theories is outlined.
\end{abstract}


\narrowtext
\section{Introduction}
\label{sec1}
Currently there is  considerable interest in the behavior of low-momentum fields in a 
high-temperature plasma environment. For a bosonic field $\phi$ the objects of study  
are the time-dependent correlation functions for the soft modes
\be
\langle\phi(t_1,\vec{p}_1)...\phi(t_n,\vec{p}_n)\rangle\;\;\;\;
|\vec{p}_i|\sim g^2 T 
\label{corfunct}\ee
and one would like to derive an effective theory that governs the 
real-time dynamics of these correlation functions for the long-wave-length modes.  
In this paper we discuss the derivation of such an effective theory for scalar field theories and outline  the extension to gauge theories.

The standard approach to deriving an effective theory for the soft modes is the introduction of an intermediate energy-scale \(\Lambda <<T\) that explicitly separates the hard modes with $\;|\vec{p}|>\Lambda$ from the soft modes with $|\vec{p}|<\Lambda$.  The hard modes are integrated out as irrelevant degrees of freedom and the effective theory of the soft modes is approximated by a classical theory \cite{bodeker,greiner,son,hu,buchmuller}. 
However there are two problems:
firstly, the results may depend on the scale \(\Lambda\) \cite{bodeker,greiner},
and, secondly, such a cut-off breaks gauge invariance \cite{bodeker}. 

We will present here a different approach, without intermediate cutoff, in which the  classical effective theory is constructed as a ``natural'' extension of static dimensional reduction to real-time.  The essence is that the zero modes of static dimensional reduction are given the role of initial conditions for classical background fields satisfying 
effective equations of motion in the environment of the quantum fluctuations which 
are taken into account perturbatively.  
This method we will call real-time dimensional reduction.

\section{Real-Time Dimensional Reduction}

Our starting point is the standard formulation of real-time  
thermal field theory in which the temporal dependence of the fields has support on the 
contour depicted in fig. 1. 
\begin{figure}
\centerline{\psfig{figure=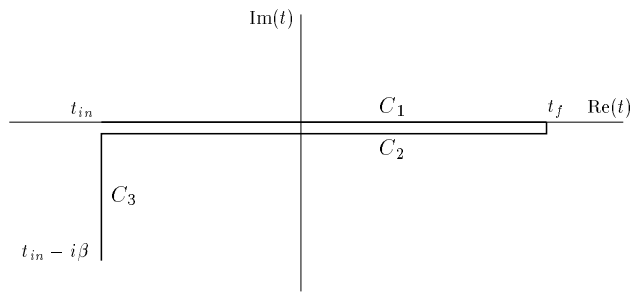,height=1.5in,angle=-90}}
\caption{The time-contour \(C\) of thermal field theory.}
\end{figure}
Equilibrium correlation functions may be obtained from the generating functional
\be
Z[j]=\int \D\phi\D\pi\exp iS[\phi,\pi]+ij\cdot\phi
\label{genfun}
\ee
with \(\phi(x)\) periodic over the entire contour \(C\). 
The dot-notation in the source term is an abbreviation for the inner product
$j\cdot\phi=\int_{C}dt d^3x j(x)\phi(x)$.
Since we are interested in real-time correlation functions, we only allow for 
sources on the real-time part of the contour $C_{12}=C_{1}\cup C_2$.
We shall keep the integration over the momenta \(\pi(x)\) explicit, since we want to end up with an effective classical theory with classical initial conditions on fields and conjugate momenta.

We now separate off the integration over the zero modes \(\Phi=\Phi(\vec{x})\)
and \(\Pi=\Pi(\vec{x})\) according to
\bea
Z[j]&=&\int \D\Phi \D\Pi\int \D\phi\D\pi
\delta\left({\cal P}\phi-\Phi\right)\delta\left({\cal P}\pi-\Pi\right) \nn\\
& &\;\;\;\;\;\;\;\;\;\;\;\;\;\;\exp iS[\phi,\pi]+i j\cdot\phi 
\label{extrzermod}\eea
where the operator \({\cal P}\) projects out the static zero mode on the Euclidean part of the contour $C$
\be
{\cal P}\phi=iT\int_{C_3}dt\phi(t,\x)
\label{projector}\ee
On the real-time part $C_{12}$ of the time-contour the 
path integration is unrestricted.  This yields the important corollary that 
local symmetries on $C_{12}$  are respected in the path integral 
(\ref{extrzermod}) which is especially useful for gauge theories.

\section{Static Dimensional Reduction}
To gain some understanding of the formal manipulation performed in 
(\ref{extrzermod}), let us consider for a moment a purely Euclidean theory with
a time-contour running straight down from 
$t_{\rm in}\) to \(t_{\rm in}-i\beta$, that is, the contour \(C\) 
with 
$t_{\rm in}=t_{\rm f}$ . 
In this case the field is periodic on $C_3$ and may be expanded in Matsubara modes
\be
\phi(t_{\rm in}-i\tau,\x)=\sum_{n}\phi_{n}(\x)e^{i\omega_{n}\tau}
\label{matsubara}
\ee
with Matsubara frequency \(\omega_{n}=2\pi n T\). The operator \({\cal P}\)
then just projects out the zero Matsubara mode 
\be
{\cal P}\phi=T\int_{0}^{\beta}d\tau \phi(t_{\rm in}-i\tau,\vec{x})=\phi_{0}(\vec{x})
=\Phi(\vec{x})
\label{zeromode}
\ee
and similarly the zero momentum mode ${\cal P} \pi=\Pi$. The path integral 
over the field \(\phi\) in
(\ref{extrzermod}) simply factorizes into an integration over the zero modes 
and one over the non-zero Matsubara modes
\be
\int\D\phi\delta\left({\cal P}\phi-\Phi\right)\rightarrow 
\int \D\phi_{n\not=0}
\label{pathint}\ee
and we end up with the effective static theory
\be
Z=\int \D\Phi \D\Pi\exp iS[\Phi,\Pi]+iW_{DR}[\Phi]
\label{stattheor}\ee
where the effective action \(W_{DR}\) is defined by
\be
\exp iW_{DR}[\Phi]=\int \D\phi_{n\not=0} \exp iS[\Phi;\phi_{n \not=0}]
\label{statdr}
\ee
with \(S[\Phi;\phi_{n\not=0}]\) the action of the non-zero Matsubara modes 
in a static background field \(\Phi\). 
For scalar fields the effective action does not depend on the static momenta, 
since momenta appear only linearly and quadratically in the action.
We recognize (\ref{stattheor}) and 
(\ref{statdr}) as the standard formulation of static dimensional reduction
\cite{landsman,jakovac,kajantie}.

The effective action $W_{DR}[\Phi]$ gives the corrections of the non-zero 
Matsubara modes to the classical action.
Since the non-zero Matsubara modes have masses of the order of the temperature,
no IR problems occur when the non-zero Matsubara modes are integrated out.
This implies that the
effective action has the property that it is in principle perturbatively calculable.

\section{Effective Action}
Returning now to the real-time case we introduce two classical fields 
\(\phi_{\rm cl},\pi_{\rm cl}\) as background fields on the entire contour
\bea
\phi(t,\x)&\rightarrow &\phi(t,\x) +\bphi(t,\x)\nn\\
\pi(t,\x)&\rightarrow &\pi(t,\x) +\bpi(t,\x) 
\eea

On the Euclidean branch of the contour the classical fields are chosen to be 
constant in time and equal to the zero modes
\be
\bphi(t,\x)=\Phi(\x) \;,\;
\bpi(t,\x)=\Pi(\x)\mbox{ on } C_3
\label{infields}
\ee
On the real-time part of the contour the classical fields will satisfy 
effective equations of motion (to be specified later on) with initial 
conditions \(\Phi,\Pi\).

Performing the shift in (\ref{extrzermod}), we can cast the generating functional in the form  
of a path integral over the initial state 
\be
Z[j]=\int D\Phi\D\Pi\exp iS_{\rm cl}+iW[\bphi,\bpi;J]+ij\cdot\bphi
\label{genfun2}\ee
with classical action \(S_{\rm cl}=S[\bphi,\bpi]\).
The quantum corrections are contained in the effective action
\bea
\exp iW[\bphi,\bpi;J]&=&\int \D\phi\D\pi\delta\left({\cal P}\phi\right)
\delta\left({\cal P}\pi\right)\nn\\
& &\exp iS[\bphi,\bpi;\phi,\pi]+iJ\cdot\phi
\label{effact}\eea
where we used that the classical fields on \(C_{3}\) are equal to the zero 
modes to simplify  the delta functions
\be
\delta\left({\cal P}(\phi+\bphi)-\Phi\right)=\delta\left({\cal P}\phi\right)
\label{delta}\ee
The action \(S[\bphi,\bpi;\phi,\pi]\) describes the quantum fields in a 
background of the classical 
fields. By definition it contains only the terms of quadratic-and-higher order in
the quantum fields \(\phi,\pi\). 
By demanding \(\pi_{\rm cl}=\partial_{t}\bphi\) on $C_{12}$ the term linear in \(\pi\)
has been made to vanish, and the linear term in \(\phi\) has been  
absorbed in the source $J=j+\delta_{\phi}S|_{\phi = \phi_{\rm cl}}$.

\section{Propagator}
We consider the example of a $\lambda\phi^4$-model and split the action of the quantum fluctuations in (\ref{effact}) into a free and interaction part: 
\be
S[\phi_{\rm cl},\pi_{\rm cl};\phi,\pi]=S_{0}[\phi,\pi]+
S_{I}[\phi_{\rm cl};\phi]
\ee
with the interactions given by
\be
S_{I}[\phi_{\rm cl};\phi]=\frac{1}{4}
\begin{array}{c}\mbox{\psfig{figure=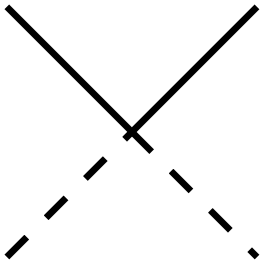,height=0.5in,angle=-90}}\end{array}
+\frac{1}{3!}
\begin{array}{c}\mbox{\psfig{figure=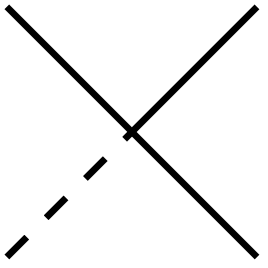,height=0.5in,angle=-90}}\end{array}
+\frac{1}{4!}
\begin{array}{c}\mbox{\psfig{figure=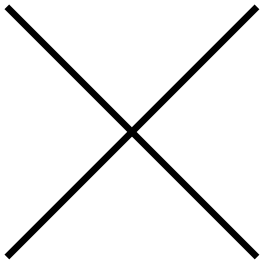,height=0.5in,angle=-90}}\end{array}
\label{intact}
\ee
The dashed lines denote the classical background fields and the 
solid lines the quantum fields.

Assuming that the effective action can be calculated perturbatively, to be checked afterwards, we have
\be
\exp iW[\bphi;J]=\exp iS_{I}[\bphi;-i\delta_{J}]\exp iW_{0}[J]
\label{pertact}\ee
with \(W_{0}[J]\) the effective action of the free theory. 
As it turns out for a scalar \(\phi^4\)-theory the effective action does not depend on the classical momentum. 
The free effective theory can be 
calculated by a Gaussian integration over the fields, but with a constraint 
on the integration enforced by the \(\delta\)-functions in (\ref{effact}). The 
result is \cite{nauta}
\be
\exp iW_{0}[J]=\exp-\half J\cdot\Delta_{C}\cdot J
\label{freeact}\ee
where the propagator of the quantum fluctuation 
\be
\Delta_C(t-t';\vec{k})=D_C(t-t';\vec{k})-S_C(t-t';\vec{k})
\label{qprop}
\ee
is equal to the thermal propagator \(D_{C}\) 
with a subtraction \(S_{C}\) due to the integration constraints. 
The thermal propagator on the contour has the standard form \cite{lebellac}
\bea
D_C(t-t';\vec{k})&=&\frac{1}{2\omega_{k}}\left[\Theta_{C}(t-t')e^{-i\omega_{k}(t-t')}\right.\nn\\
& &\;\;\;\;\;\;\,+\Theta_{C}(t'-t)e^{i\omega_{k}(t-t')}\nn\\
&+ &\left.n(\omega_{k})\left(e^{-i\omega_{k}(t-t')}+e^{i\omega_{k}(t-t')}\right)
\right]
\label{therprop}
\eea
with energy \(\omega_{k}^2=\vec{k}^2+m^2\) and Bose-Einstein distribution 
\(n(\omega)=(e^{\beta\omega}-1)^{-1}\),
whereas the subtraction on the various parts of the contour is given by the expressions
\be
S_C(t-t',\vec{k})=\left\{
\begin{array}{ll}
\frac{T}{\omega_{k}^2}\cos\omega_k(t-t')& t,t'\in C_{12}\\
 \\
\frac{T}{\omega_{k}^2}\cos\omega_k(t_{in}-t')& t\in C_{3}, t'\in C_{12}\\
 \\
\frac{T}{\omega_{k}^2}\cos\omega_k(t- t_{in})&t\in C_{12}, t'\in C_{3}\\
 \\
\frac{T}{\omega_{k}^2}&t,t'\in C_{3}
\end{array}
\right.
\label{clasprop}
\ee
On \(C_{12}\) we recognize \(S_{C}\) as the classical propagator
and on \(C_3\) as the zero-mode propagator. The two contributions connecting the vertical branch \(C_3\)
of the contour with the real-time branch \(C_{12}\) depend on the initial time \(t_{\rm in}\). In thermal field theory it is shown that in the limit $t_{in}
\rightarrow -\infty $, the two horizontal real-time contours $C_{12}$
decouple from the Euclidean branch $C_3$, provided that one introduces an infinitesimal damping coefficient \cite{lebellac}. 
However, we have to keep \(t_{\rm in}\)
finite. Then obviously the contributions of $C_3$ and $C_{12}$ do not separate
and  the system retains a memory of the initial state. How this works 
out in the equations of motion is discussed in Appendix A.

Let us consider the IR limit \(\omega_{k}/T\rightarrow 0\)
of the propagator of the quantum fields (\ref{qprop}). 
The dominant contribution to the 
thermal propagator comes from the thermal part of (\ref{therprop}), since 
\(n(\omega_{k})\rightarrow T/\omega_{k}\). Therefore, in the high-temperature limit the thermal propagator reduces to  
\be 
D_C(t_1-t_2;\vec{k})\rightarrow S_C(t_1-t_2;\vec{k})\;\;
(\omega_{k}\rightarrow 0)
\label{IRlimit}\ee
when we ignore terms
\(1/\omega_{k}\) compared to \(T/\omega_{k}^2\).

In particular on \(C_{3}\) the propagator of the quantum fluctuations is
just the propagator of static dimensional reduction, namely the propagator 
of the non-static Matsubara modes
 \be
\Delta_C(-i\tau,\vec{k})=T\sum_{n\not=0}
\frac{e^{i\omega_{n}\tau}}{\omega_{n}^2+\omega_{k}^2}
\label{nonstatprop}\ee
Consequently, the effective action will contain as a subset also  all diagrams of the Euclidean dimensionally reduced action.

From (\ref{IRlimit}) and (\ref{nonstatprop}) we see that 
the propagator of the quantum fluctuations is 
better IR-behaved than the thermal propagator. Hence the important 
conclusion is that a perturbative evaluation of the effective action is
possible without encountering the severe IR divergences of the full thermal
theory.
In other words: the proposed splitting of the classical and quantum degrees 
of freedom isolates the IR sensitive part of the theory in the classical 
part of the theory. 

\section{Equations of motion}

Let us return to the generating functional (\ref{genfun2}). The classical and effective action in the integrand depend on the classical fields $\bphi,\bpi$ which were defined both on \(C_{12}\) and \(C_{3}\). The contributions of these two separate contour segments 
have a completely different interpretation. This difference can be made more explicit by choosing the classical background
field \(\bphi\) equal on the upper and lower real-time branch
\be
\bphi(t_1,\vec{x})=\bphi(t_2,\vec{x}) \;\; t_1=t\in C_{1},\; 
t_2=t \in C_{2}
\label{phieqc1c2}\ee
Recalling also (\ref{infields}), we see that the two contributions of the classical action on the upper and lower time-branch cancel, and that we are left with 
\(iS[\bphi,\bpi]=-\beta H[\Phi,\Pi]\), which is the classical thermal weight in terms of the Hamiltonian.  In a similar fashion, the effective action may be split into two parts
\be
W[\bphi;J]=W_{IF}[\bphi;J]+W_{DR}[\Phi]
\ee
with an influence action \(W_{IF}\) that vanishes at zero source and 
\(W_{DR}\) the effective action of static dimensional reduction (\ref{statdr}),
which depends solely on the initial field \(\Phi\).

The generating functional may then be written 
\bea
Z[j]&=&\int\D\Phi\D\Pi\exp \left(-\beta H[\Phi,\Pi]+iW_{DR}[\Phi]\right)\nn\\
& &\;\;\;\;\;\;\;\;\exp\left(iW_{IF}[\bphi;J]+ij\cdot\bphi\right)
\label{genfun3}
\eea
where we have separated the exponent into the effective thermal weight factor 
and  a second factor which is the effective source term for the real-time 
correlation functions.

The expression above is still quite general as no approximation has been 
made. All complications of the interaction between hard and soft modes are 
contained in the influence functional. However, this functional may be 
calculated perturbatively and we are only looking for an effective theory 
that contains the leading high-temperature behavior.  Moreover, we still 
have not made a unique choice for the classical field \(\bphi\) on \(C_{12}\).
An obvious choice is to equate the mean classical field with the full mean 
field including quantum fluctuations 
\be
\left. \delta_j W_{IF}[\bphi;J]\right|_{j=0} =0
\ee
This may be translated into the equation of motion
\be
\left. \delta_\phi \Gamma[\phi] \right|_{\phi=\bphi} =0
\label{eqmotion}
\ee
with $\Gamma[\phi]$ the Legendre transform of $W_{IF}[\bphi;J]$ with $\bphi=0$ 
on $C_{12}$ \cite{raifeartaigh}. If we now confine ourselves to the leading 
contribution, this reduces to 
\be
\delta_{\phi}\left.\Gamma_{HTL}[\phi]\right|_{\phi=\bphi}=0
\label{HTLeqmot}\ee
with \(\Gamma_{HTL}[\phi]\) the hard thermal loop (HTL) effective action
in the low-momentum limit. 
It is noteworthy to mention that large contributions from soft modes on internal lines
of Feynman diagrams that arise in thermal theories are not present in the effective action, since the IR-sensitive part of the thermal propagator is subtracted in (\ref{qprop}). As a consequence, the leading contributions do come solely from the hard thermal loops.

Since correlation functions are evaluated at zero-source, the leading contribution to statistical correlation functions is given by the classical 
contribution, that is, the influence function in (\ref{genfun3}) can be 
ignored in this calculation. 
However, if we calculate, for instance, the 
expectation value of the commutator of fields, the 
classical contribution is zero (classical fields commute). Then the leading 
HTL contribution has to be obtained from the influence action by source 
differentiation. 

In scalar \(\phi^4\)-theory, the only hard thermal loop contribution is the 
tadpole contribution to the thermal mass 
$m^2_{\rm th}=m^2+\lambda T^2/24$. When calculating this contribution we 
encounter a linear divergence,  that may be set to zero if we use the 
dimensional regularization scheme. However, as a general rule, UV-divergences 
do not disappear but
act as counter terms for the same divergences occurring in the classical 
theory.  This general feature is easily understood from the fact that the 
full quantum theory is renormalizable. Hence, all classical divergences must 
be exactly neutralized by corresponding counter terms from the quantum 
corrections.   

Because we only have to take into account the thermal mass, the HTL equations 
of motion (\ref{HTLeqmot}) for 
\(\lambda \phi^4\)-theory simply read
\be
(\partial_{t}^2-\vec{\nabla}^2+m_{\rm th}^2) \bphi(t,\vec{x})
+\frac{1}{3!}\lambda \bphi^3(t,\vec{x})=0
\label{expleqmot}\ee
with initial conditions
\be
\bphi(t_{\rm in},\vec{x})=\Phi(\vec{x})\;;\;
\bpi(t_{\rm in},\vec{x})=\Pi(\vec{x})
\ee
So in combination with (\ref{genfun3}) we have obtained a classical 
statistical theory, with the classical field satisfying the equation of 
motion (\ref{expleqmot}) and  
thermal averaging over the initial 
conditions. The effect of the initial conditions on the equation of motion 
is studied further in appendix A.

As shown above, the perturbative approach to the effective classical theory, amounts to  
a resummation of the thermal mass into the classical propagator, while leaving the quantum propagator unmodified. This is similar to the resummation scheme of \cite{espinosa,leupold} for static quantities, where the thermal mass is resummed in the 
zero-mode propagator.

\section{SU($N$) gauge theory }    \label{sect:7}
The method developed in this paper to study the IR-behavior of time-dependent fields,
has an obvious application in non-Abelian gauge theories, which have a 
non-trivial IR structure \cite{linde,gross,arnold,moore}. The reasoning can 
be extended almost unchanged to derive an 
effective classical statistical theory for gauge fields\cite{nauta2}. The 
final result for the generating functional (in the Coulomb gauge) is
\bea
Z[j]&=&\int \D{\cal A}^{\mu}\D{\cal E}^{l}\D{\cal C}\D\bar{{\cal C}}\nn\\
& &\;\;\;\;\;\;\exp \left(-\beta H[{\cal A},{\cal E}]+
iS_{\rm gh}[{\cal C},\bar{{\cal C}},{\cal A}]+
iW_{DR}\right)\nn\\
& &\;\;\;\;\;\;\exp\left(i W_{IF}[A_{\rm cl},E_{\rm cl};J]+ij\cdot A_{\rm cl}\right)
\label{genfungauge}\eea
with ${\cal A, E}$ the static gauge fields, \({\cal C},\bar{{\cal C}}\) the ghost zero modes,  and \(S_{\rm gh}\) the ghost action. We have extracted the effective classical theory that gives the 
leading-order contributions in the low-momentum limit.
Subleading corrections in \(g\) or \(|\vec{p}|/T\) 
are given by the quantum corrections contained in the effective actions
\(W_{IF},W_{DR}\). Although subleading these 
contributions may be important to provide counter terms for the classical 
divergences, as in the static case \cite{jakovac}.

The classical fields have to be determined from the HTL equation of motion, 
but calculated with the gauge propagators equivalent to (\ref{qprop})
\be
[D^{\nu}_{\rm cl},F^{\rm cl}_{\nu\mu}](x)=3\omega_{p}^2\int\frac{d\Omega}{4\pi}
v_{\mu}\int_{-\infty}^{t}dt'U_{\rm cl}(x,x')
v_{\nu}F^{0\nu}_{\rm cl}(x')
\label{effeqsun}
\ee
with $\omega_{p}^2=N g^2 T^2/6$ the plasmon frequency. 
The angular integration is over the direction of \(\vec{v}\), 
\(|\vec{v}|=1\).  Furthermore, \(x'=(t',\vec{x}-\vec{v}(t-t'))\)
and 
\(U_{\rm cl}(x,x')=\mbox{P}\exp\left( -ig\int_{\gamma} dz_{\mu}
A^{\mu}_{\rm cl}(z)\right)\) , 
with \(\gamma\) a straight line from 
\(x\) to \(x'\), is the parallel transporter.
The initial conditions are
\be
A_{\rm cl}^{\mu}(t_{\rm in},\vec{x})={\cal A}^{\mu}(\vec{x})\;;\;
F^{0l}_{\rm cl}(t_{\rm in},\vec{x})={\cal E}^l(\vec{x})
\ee
The field \(A_{\rm cl}\)
 at times prior to \(t_{\rm in}\) on the r.h.s. of (\ref{effeqsun})
 should be taken constant in 
time and equal to the initial condition. This prescription takes into account 
the correlations between the initial conditions and the fields at later times,
as explained in Appendix A.

Equation (\ref{effeqsun}) is the well known HTL equation of motion, 
derived in a kinetic approach by Blazoit and Iancu \cite{blazoit}, 
where asymptotic initial 
conditions are considered. We have initial conditions at a finite time 
\(t_{\rm in}\) over which a statistical average has to be taken.
This implies that the physics at the scale \(g^2 T\) is still present 
in the effective theory.

The SU($N$) effective theory has two obvious applications. The study of the 
IR sector of the theory, for instance by deriving an effective theory for 
the non-perturbative modes, that may be studied on the lattice \cite{iancu}. 
And secondly to study the effect of IR effects on for instance plasmon 
properties, this approach may go beyond the phenemonological introduction 
of a magnetic mass to account for non-perturbative effects in the IR sector 
of the theory.

\section{Conclusions}

We have shown that the generating functional of a weakly coupled field theory
at high temperatures may be formulated as an effective classical statistical
theory. The quantum corrections are calculable in perturbation theory and 
enter the thermal weight and source term, see (\ref{genfun3}) and 
(\ref{genfungauge}).

For a low-momentum process containing no classical UV divergences the quantum 
corrections in the influence action give sub-leading contributions to the 
leading order classical result. If the process contains UV divergences
(that is, it is sensitive to the physics at high momenta in the classical 
theory) the influence action and the  dimensionally reduced static action 
provide the counter terms to render the result finite, since (\ref{genfun3})
and (\ref{genfungauge}) are still equivalent to the full theory.
For scalar \(\lambda\phi^4\)-theory it was shown in \cite{aarts1,aarts2} 
that local counter terms are sufficient to render the classical theory finite.
In general (gauge theories) non-local counter terms will be needed 
\cite{arnold}.

We have noted that the extraction of the zero modes 
(\ref{extrzermod}) and the subsequent manipulations do not break local 
symmetries on \(C_{12}\). Especially for gauge theories this is a useful 
property. Although, we have broken 
gauge symmetry (in section \ref{sect:7} we have chosen the Coulomb gauge) to perturbatively evaluate the effective action, the remaining BRST invariance is preserved. This means that the action 
\(\Gamma[\phi]\) in the equations of motion (\ref{eqmotion}) for gauge theories 
satisfies the Slavnov-Taylor identities. In the HTL approximation this is upgraded to gauge invariance \cite{kobes}.

As mentioned already in the Introduction
often an intermediate momentum scale \(\Lambda\)
is introduced that separates between the classical and the quantum modes.
In \(\lambda\phi^4\) this scale is chosen such that \(m_{\rm th}<<\Lambda<<T\) 
\cite{greiner,buchmuller}. In our formalism such a scale can 
be introduced by extracting only zero modes with momenta \(|\vec{k}|<\Lambda\)
in (\ref{extrzermod}). The subtraction for the propagator in (\ref{qprop})
is then restricted to momenta \(|\vec{k}|<\Lambda\)
\be
S(t-t';\vec{k})\rightarrow \Theta(\Lambda-|\vec{k}|)S(t-t';\vec{k})
\ee
This way the actions \(W_{DR},W_{IF}\) and the equations of motion
become \(\Lambda\)-dependent. These \(\Lambda\)-dependent terms may be 
interpreted as counter terms for the \(\Lambda\)-dependence of the classical 
theory; of course, the full generating functional should be independent of the 
choice of \(\Lambda\).

The dependence on the cut-off \(\Lambda\) of the classical theory
can be inferred from its UV behavior. Since \(\Lambda\) is larger than the 
typical classical energy-scale the divergences in the theory without cut-off
get replaced by powers of 
\(\Lambda/m_{\rm th}\)
or \(\log \Lambda/m_{\rm th}\) (in \(\lambda \phi^4\)-theory).
So the problem of introducing counter terms to cancel 
all \(\Lambda\)-dependences, is the same problem of introducing 
counter terms for the divergences as \(\Lambda\rightarrow \infty\). 
It has been argued \cite{iancu} that in a leading order
calculation of IR-sensitive quantities
the non-HTL contributions that depend 
on \(\Lambda\) may be neglected for \(\Lambda\) in the range 
\(gT<<\Lambda<<T\) (in gauge theories). 
Since these non-HTL contributions are suppressed 
by powers of the coupling constant or the classical energy scale over the 
temperature.

The introduction of such a cut-off scale may be convenient to make a 
derivative expansion for the non-local terms in the effective action and the 
equations of motion.
However we would like to emphasize that for the derivation of the classical 
theory no intermediate scale \(\Lambda\) is necessary. As in static 
dimensional reduction the relevant parameter is the classical energy-scale 
over the temperature, not \(\beta\Lambda\) but \(\beta m_{\rm th}\)
in \(\lambda \phi^4\)-theory.

\appendix
\section{Initial correlations}
The effective action \(\Gamma[\phi]\) in (\ref{eqmotion}) depends on the 
zero mode \(\Phi\). Therefore one may expect that in the equations of motion 
the zero mode will also occur, describing the correlations between the 
initial conditions and the classical field at later times. 
As an example let us consider the linear equations of motion
\be
(\partial_{t}^2+\omega_{k}^2)\bphi(t,\vec{k})+
\int_{C} dt'\Sigma(t-t',\vec{k})\bphi(t',\vec{k})=0
\label{lineqmotion}
\ee
with the time integral over the entire contour.
Since the classical field is equal on the forward and backward time-branches, 
the real-time part of the self-energy entering the equations of motion is 
given by the retarded self energy
\bea
\int_{C} dt'\Sigma(t-t',\vec{k})\bphi(t',\vec{k})&=&
\int_{t_{\rm in}}^{\infty}dt' \Sigma_{R}(t-t',\vec{k})\bphi(t',\vec{k})
\nn\\
&+&\int_{C_{3}}dt' \Sigma(t-t',\vec{k})\Phi(\vec{k})
\eea 
The second term on the r.h.s. gives the correlations with the initial field.
Since \(t_{\rm in}\) is finite this term cannot be dropped. However, we may rewrite it in a more convenient form.
For this we deform the contour \(C_{3}\) to first run back to some earlier
time \(t^{*}_{\rm in}\) then down the Euclidean path \(C_{3}^{*}\)
to \(t_{\rm in}^{*}-i\beta\)
and finally straight back to \(t_{\rm in}-i\beta\). Using the assumption of 
analyticity of the self-energy in the complex time plane 
and periodicity, we arrive at
\bea
\int_{C_{3}}dt' \Sigma(t-t',\vec{k})&=&
\int_{t^{*}_{\rm in}}^{t_{\rm in}}dt' \Sigma_{R}(t-t',\vec{k})\nn\\
&+&\int_{C^{*}_{3}}dt' \Sigma(t-t',\vec{k})
\label{decoup}
\eea
Introducing an infinitesimal damping rate, the decoupling theorem of thermal field theory states that 
the second term at the r.h.s. of  (\ref{decoup}) may be dropped in the limit 
\(t^{*}_{\rm in}\rightarrow -\infty\) \cite{mabilat}. 

Returning to the equation of motion (\ref{lineqmotion}), we find
\be
(\partial_{t}^2+\omega_{k}^2)\bphi(t,\vec{k})+
\int_{-\infty}^{\infty}dt' \Sigma_{R}(t-t',\vec{k})\bphi(t',\vec{k})
=0
\label{reseqmot}
\ee
where it should be understood that the classical field before \(t_{\rm in}\)
appearing in the memory kernel equals the initial value: 
\(\bphi(t,\vec{k})=\Phi(\vec{k})\) for \(t<t_{\rm in}\).
Physically this means that the system is frozen in the initial state from the infinite past till the initial time  \(t_{\rm in}\) where a disturbance is applied.
This result can easily be generalized to non-linear equations of motion with 
higher-point vertex functions.

\end{document}